\documentclass[a4paper]{article}
\usepackage{INTERSPEECH2021}

\usepackage{amsmath,graphicx}

\usepackage{subcaption}
\usepackage{bbold}

\usepackage{blindtext}
\usepackage{color}
\definecolor{mypink1}{rgb}{0.858, 0.188, 0.478}
\definecolor{mycolor3}{rgb}{0, 0.5, 0.5}
\definecolor{mycolor2}{rgb}{0, 0, 0.9}
\definecolor{mygreen}{rgb}{0, 1, 0}
\definecolor{mymaroon}{rgb}{0.5, 0, 0}

\title{Acted vs. Improvised: Domain Adaptation for Elicitation Approaches in Audio-Visual Emotion Recognition}
\name{Haoqi Li$^{1,3,*}$ \thanks{* Work performed while interning at Amazon.}, Yelin Kim$^2$, Cheng-Hao Kuo$^2$, Shrikanth Narayanan$^3$}
\address{
  $^1$Amazon AWS AI 
  $^2$Amazon Lab126\\
  $^3$Dept. of Electrical and Computer Engineering, University of Southern California, USA}
\email{\{haoqili, kimyelin, chkuo\}@amazon.com, shri@sipi.usc.edu}

\begin{document}
\ninept
\maketitle
\begin{abstract}
Key challenges in developing generalized automatic emotion recognition systems include scarcity of labeled data and lack of gold-standard references. 
Even for the cues that are labeled as the same emotion category, the variability of associated expressions can be high depending on the elicitation context e.g., emotion elicited during improvised conversations vs. acted sessions with predefined scripts. 
In this work, we regard the emotion elicitation approach as domain knowledge, and explore domain transfer learning techniques on emotional utterances collected under different emotion elicitation approaches, particularly with limited labeled target samples. 
Our emotion recognition model combines the gradient reversal technique with an entropy loss function as well as the softlabel loss, and the experiment results show that domain transfer learning methods can be employed to alleviate the domain mismatch between different elicitation approaches. 
Our work provides new insights into emotion data collection, particularly the impact of its elicitation strategies, and the importance of domain adaptation in emotion recognition aiming for generalized systems.

\end{abstract}
\vspace{-0.2cm}
\section{Introduction}
\label{sec:intro}
Human emotion expressions can provide clues about individual needs, preferences, and attitudes;  emotion-related cues can be used to create more natural and human-centered interactive technology.
Recent deep learning techniques have shown promising improvements in automatic emotion recognition \cite{schuller2018speech, tzirakis2017end}; however, building an accurate emotion recognition system is still challenging due to the shortage of labeled data, and the lack of gold-standard references.
Moreover, the performance of emotion recognition system is often sensitive to the variations across different domains.
These domain variations can be potentially caused by many factors such as the speaker characteristics \cite{zhan1997speaker} and recording conditions \cite{schuller2006emotion}, emotion elicitation approaches \cite{proc_carlos:improvisationelicitationiemocap:interspeech08, busso2016msp, petrantonakis2011novel, parada2020demos} or even socio-cultural context \cite{lim2016cultural, zhao2019adversarial}.
Hence, emotion recognition systems that are only evaluated within a single domain condition may not be robust to these variations. 

In automated emotion recognition (ER) \cite{wu2014survey, feng2020review, abbaschian2021deep}, transfer learning has been employed to alleviate the domain mismatch in the cross-corpora \cite{milner2019cross, song2019transfer, latif2018transfer, gideon2019improving, lee2021domain}, cross-language \cite{latif2018transfer,chiou2014speech,latif2019unsupervised}, cross-speaker \cite{li2020speaker, yin2020speaker} and cross-modality \cite{mariooryad2013exploring, chao2018generating} scenarios.
Besides those above, emotion elicitation approaches can be another important mismatch factor in ER.
Existing works \cite{proc_carlos:improvisationelicitationiemocap:interspeech08, busso2016msp} already show utterances with the same emotion label can have different expressed behaviors in natural conversations, scripted acted talks and improvised conversations. 
However, related work of quantitative inferential capacity across different emotion elicitation approaches is still very limited.

The elicitation approaches for most emotion recognition corpora can be divided into two categories: (1) read-speech with fixed lexical scripts and (2) spontaneous speech collected from improvised interaction sessions.
In this work, we analyze the domain mismatch and inferential capacity across utterances generated from different \textit{emotion elicitation approaches}.
We explore two hypotheses:
\textbf{H1}: Improvised utterances as a source domain can help the emotion recognition of the scripted utterances and \textbf{H2}: Using scripted utterances as a source domain can improve the emotion recognition of the improvised utterances.

\textbf{H1} is motivated by real human-agent interactions. Many widely-deployed speech agents (e.g. at home, cars etc.) are likely used in specific target domains (e.g. “weather”, “music” etc.), in which most commands have relatively fixed and specialized content. 
Thus, the potential of leveraging a larger corpus with free content (e.g. public emotional utterances collected from Youtube) for target domain adaptation is explored under source-target setting in \textbf{H1}.
Currently, a large number of emotional utterances can be easily collected from online resources, such as online videos \cite{morency2011towards} and call recordings \cite{lee2005toward, devillers2006real}, however, copyright issues, privacy, and lack of controlled recording settings (background noise, etc.) might cause serious issues in commercial products or usage.
Thus, \textbf{H2} is motivated in that the many available curated emotion datasets are collected in carefully designed settings with controlled conditions, including with user consent, and leveraging these data for improving performances on natural sessions can be highly impactful.

To alleviate the domain mismatch, traditional transfer learning approaches, such as fine-tuning on target domain, often require hundreds of labeled samples, which can be challenging in the ER area considering the scarcity and cost of labeled data. 
Recently, domain adversarial training \cite{ganin2016domain} has been employed in ER to remove domain information \cite{li2020speaker,abdelwahab2018domain, peri2020empirical}. 
In our work, to verify the two proposed hypotheses above, we build an emotion recognition model with the combination of adversarial iterative training strategy and softlabel loss from work \cite{li2020speaker} and \cite{tzeng2015simultaneous}.
The adversarial iterative training strategy, on the one hand, is to maximize the capture of emotion-related information; on the other hand, is to remove domain information (the difference between various emotion elicitation approaches).
The introduction of softlabel loss ensures that domain transfer learning can consider both domains and emotion categories simultaneously. This is motivated by constraints in real-world use settings, where it is not feasible to have enough target domain labeled data to conduct supervised tuning for each customer or each new usage scenario.
\vspace{-0.2cm}

\begin{figure*}[t]
\centering
\includegraphics[width=0.7\textwidth]{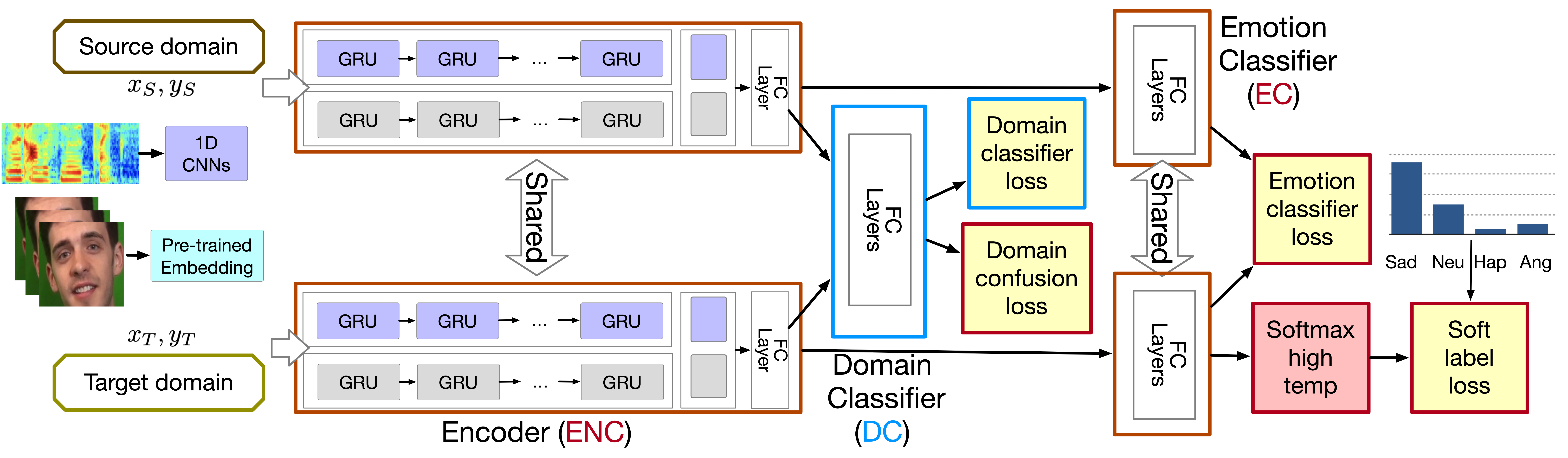}
\caption{Our model includes the representation encoder (ENC), one domain classifier (DC), and one emotion classifier (EC). We use a domain classifier loss to optimize the DC and use domain confusion loss, emotion classifier loss and softlabel loss to remove the domain information and optimize the emotion classification accuracy.}
\label{fig:model}
\vspace{-0.4cm}
\end{figure*}

\section{Methodology}
Our work aims to (1) investigate domain adaptation across utterances obtained through different emotion elicitation approaches and (2) apply domain adversarial learning and softlabel loss with limited target samples.
The goal of adversarial iterative training strategy is to predict emotion class labels while simultaneously finding representations that make the domains as similar as possible. Technically, a max-entropy adversarial network model from \cite{li2020speaker} is utilized by optimizing over a loss which includes both classification error on the labeled samples as well as a combined gradient reversal technique with an entropy loss function to make the domains indistinguishable. 

However, while maximizing domain confusion pulls the marginal distributions of the domains together, it doesn't necessarily align the different emotion categories of the target with those in the source. 
The correlations across different emotions are related to the properties of human emotion expression and perception. For example, from the acoustic expression aspect, happiness and anger often result in a similar higher pitch. In addition, in emotion perception, sadness can be easily confused with neutral states regardless of utterance type as shown in \cite{busso2016msp}. 
Motivated by work \cite{tzeng2015simultaneous}, besides a single emotion label, we also employ softlabel loss to perform the emotion category adaptation. We explicitly transfer the similarity of emotion categories from the source to the target and further optimize our representation to produce the same structure in the target domain using a few target labeled examples as reference points.

\vspace{-0.2cm}
\subsection{Max-Entropy Adversarial Network}
Our proposed model takes acoustic and visual features as inputs, and contains three modules: the representation encoder (ENC), the domain classifier (DC), and the emotion classifier (EC) as shown in Figure~\ref{fig:model}.

For the acoustic ENC, we use stacked 1D convolutional layers and followed by sequential GRU layers \cite{cho2014learning} to obtain the acoustic representation for each utterance.
For the visual ENC, we extract the intermediate embedding from a pre-trained inception-ResNet \cite{szegedy2017inception} model on a face recognition task, then use this embedding as the input feature to another GRU model to generate the visual representation. 
Then, we concatenate the acoustic and visual representations and add fully connected layers to generate the emotion representation.
This fixed dimension representation, as the output of the ENC, is further connected to two classification modules: the emotion classifier (EC) and the domain classifier (DC).
We name the parameters of each component as $\theta_{enc}$, $\theta_{EC}$ and $\theta_{DC}$, respectively. 

The model takes as input the labeled source data $\{x_S, y_S\}$, and a few target labeled data $\{x_T, y_T\}$. The goal is to train an emotion category classifier that operates on the emotion representation $enc(x)$, and can correctly classify the target samples during the evaluation.
The training strategy has the iterative update process: On the one hand, we attempt to optimize the DC to estimate the domain information; on the other hand, we attempt to correctly estimate the emotion label and remove the domain information for ENC output. 

\vspace{-0.3cm}
\subsubsection{Training of DC}

For a given representation $v$, the elicitation domain classifier (DC) can be regarded as a ``discriminator", which is trained to distinguish the domain identity. Mathematically, this can be done by optimizing a cross entropy objective function as follow:
\begin{equation}\label{eq:ce_dc}
\resizebox{.913\hsize}{!}{
$\min_{\theta_{DC}} L_D (x_S, x_T, \theta_{enc}; \theta_{DC}) =  
-\sum_{d} \mathbb{1} [y_D = d] log P(d|enc(x))$
}
\end{equation}

where $d$ is the domain label for sample $x$, and $P(d|enc(x))$ is the softmax output of domain classifier. In this training step, only the parameters of DC, i.e. $\theta_{DC}$, are optimized.

\vspace{-0.2cm}
\subsubsection{Training of ENC and EC}

For the training of ENC and EC, we need to optimize the model to improve the emotion classification accuracy. 
Simultaneously, we need to optimize ENC to increase the uncertainty or randomness of DC’s outputs. Mathematically, we maximize
the entropy value of DC’s output \cite{li2020speaker} to promote the equal likelihood for all domains.
\begin{equation}\label{eq:conf_dc}
\resizebox{.99\hsize}{!}{
$\max_{\theta_{enc}} L_{conf} (x_S, x_T, \theta_{DC}; \theta_{enc}) =  
-\sum_{d} P(d|enc(x)) log P(d|enc(x))$
}
\end{equation}
where $P(d|enc(x))$ is the softmax output of domain classifier, $d$ refers to two emotion elicitation approaches considered in our work (acted/scripted vs. natural/spontaneous). Only the parameters of ENC are optimized based on this loss function.

In addition, the performance of the emotion classifier is optimized by minimizing cross entropy loss from EC's output:
\begin{equation}\label{eq:emo_cl}
\min_{\theta_{enc}, \theta_{EC}} L_{emo} (x, y, \theta_{enc}, \theta_{EC}) =  
-\sum_{k} \mathbb{1} [y = k] log P_k
\end{equation}

where the $P$ is the softmax output of emotion classifier and $k$ is the emotion label of the training sample $(x, y) \in (x_S,y_S)$ or $(x_T,y_T)$.

To combine these two objective functions together, we flip the sign of $L_{conf}$ to do a gradient reversal and minimize the weighted overall loss sums:
\begin{equation}
\label{eq:total}
\begin{aligned}
L_{total} (x_S, y_S, x_T, y_T, \theta_{DC}; \theta_{enc}, \theta_{EC}) = \\
L_{emo} (x, y; \theta_{enc}, \theta_{EC}) - \lambda_{conf} L_{conf} (x_s, x_T, \theta_{DC}; \theta_{enc}) 
\end{aligned}
\end{equation}


We notice the loss function  (\ref{eq:ce_dc}) and (\ref{eq:conf_dc}) are in direct opposition to one another: learning a good domain classifier means the ENC output is not domain invariant, and obtaining a domain invariant ENC output can lead to a lower performance of DC. Thus, we perform the iterative update for losses (\ref{eq:ce_dc}) and (\ref{eq:total}) given the parameters from the previous iteration rather than globally optimizing all the parameters.

\vspace{-0.2cm}
\subsection{Softlabel loss}

Although the adversarial domain confusion training tries to align the marginal distribution across different domains, the alignment of classes between each domain is not guaranteed.
Thus, we further use the softlabel loss to align the source and target emotion classes, in which we try to ensure the relationship between emotion categories is preserved across source and target. This relationship originates from the human expression regardless of the elicitation approaches.

The softlabel is generated by averaging the per-category activation of source training examples using the source model. 
Thus, if in total we have $K$ emotion classes, for each emotion category $k$, we have one softlabel $l^{(k)}$ with $K$ dimensions to depict relative similarity across all emotion categories. 
The softlabel loss is defined as the cross-entropy loss between the soft activation of a target sample and the soft label corresponding to that emotion category of the sample.
\vspace{-0.1cm}
\begin{equation}\label{eq:soft}
L_{soft} (x_T, y_T, \theta_{enc}, \theta_{EC}) =  
-\sum_{i} l_i^{(y_T)}log P_i
\end{equation}
\vspace{-0.35cm}

where $ P_i$ is the softmax output of emotion classifier with a high temperature of $\tau$ to make sure related emotion classes have enough probability mass to calculate the softlabel loss.

Lastly, we include this softlabel loss (\ref{eq:soft}) to the total loss (\ref{eq:total}) to formulate the updated total loss during the training of ENC and EC as follows:
\vspace{-0.1cm}
\begin{equation}
\label{eq:total_new}
\begin{small}
\begin{aligned}
L_{total} (x_S, y_S, x_T, y_T, \theta_{DC}; \theta_{enc}, \theta_{EC}) = \\
L_{emo} (x, y; \theta_{enc}, \theta_{EC}) - \lambda_{conf} L_{conf} (x_s, x_T, \theta_{DC}; \theta_{enc}) \\
+ \lambda_{soft} L_{soft} (x_T, y_T; \theta_{DC}, \theta_{enc})
\end{aligned}
\end{small}
\end{equation}
\vspace{-0.4cm}

\begin{table}[t]
\centering
\scalebox{0.85}{
\begin{tabular}{cl}
\hline
\textbf{Utterance category}  & \textbf{Description}    \\ \hline
Natural Interaction (NI)   & {Recording natural interaction during the breaks}                                                     \\  \hline
 
Other-improvised (OI)   &  \begin{tabular}[c]{@{}l@{}}All the actors’ turns during the improvisation \\  sessions (not just the target sentences)\end{tabular}            \\ \hline
 
Target-improvised (TI)      & \begin{tabular}[c]{@{}l@{}}Sentences with fixed lexical content during \\  improvised dialog that expresses target emotions\end{tabular} \\ \hline
 
Target-read (TR)  & \begin{tabular}[c]{@{}l@{}}Actors read the sentences with fixed \\ lexical content portraying the four target emotions\end{tabular}            \\ \hline
 
Whole MSP-IMPROV & {Combination of all utterances above}  \\\hline                                                       
\end{tabular}
}
\caption{Emotional utterance type descriptions per emotional class in MSP-Improv dataset} \label{tab:msp_utt}
\vspace{-0.4cm}
\end{table}

\section{Data and Feature Extraction}
The MSP-IMPROV \cite{busso2016msp} corpus is a multi-modality emotional database, which is designed to have a controlled data collection process over lexical content and emotion while also promoting naturalness in the recordings. 
It includes four types of utterances under different emotion elicitation scenarios as shown in Tables~\ref{tab:msp_utt} and  \ref{tab:num_utt}. Within MSP-IMPROV, utterances are classified as four types based on different elicitation scenarios: Natural Interaction (NI), Other-Improvised (OI), Target-Improvised (TI) and Target-Read (TR). To the best of our knowledge, currently, the MSP-IMPROV is the only dataset that was designed with such different emotion elicitation strategies for audio-visual emotion recognition. In addition, the “emotion elicitation approach” category label was provided for each utterance.

\begin{table}[t]
\centering
\scalebox{0.87}{
\begin{tabular}{cccccc}
\hline
Utterance category       & Angry    & Sad      & Happy     & Neutral   & Total     \\ \hline
NI &  38  &  66  &  1372 &  1164 &  2640 \\ \hline
OI    &  470 &  633 &  1048 &  1789 &  3940 \\ \hline
TI   &  115 &  106 &  136  &  283  &  640  \\ \hline
TR         &  169 &  80  &  88   &  241  &  578  \\ \hline
\end{tabular}
}
\caption{Number of utterances per emotional class in MSP-Improv} \label{tab:num_utt}
\vspace{-0.4cm}
\end{table}

For each utterance, we extract 40 dimensions of Mel-filter bank coefficients and energy as acoustic features. For visual features, we use an embedding of 512 dimensions from a pre-trained face recognition model, which has a similar structure of inception-ResNet \cite{szegedy2017inception}. The extracted embedding captures the facial information and is used for emotion recognition. 

\vspace{-0.2cm}
\section{Experiment setup}

We use utterances combined from different categories as source and target domain to verify our hypotheses, and further investigate the possibility of domain transfer learning with limited samples of target domain.
In terms of the utterance content, TR and TI contain the designed scripts,  while NI and OI are mostly from natural interactions and improvisation sessions. To verify our proposed hypotheses, we design the source and target split with the following combinations:

\begin{enumerate}
    \item \textit{Using NI and OI as source domain}\\
    This split is used to test hypothesis (\textbf{H1}).
    Under this combination, source domain mainly contains utterances from natural or improvised conversations while for the target domain, lexical content of the utterances is fixed.
    \item \textit{Using TR and TI as source domain}\\
    This split is used to test hypothesis (\textbf{H2}).
    Under this setting, the lexical content is fixed in the source domain while in the target domain, the lexical content can be variable.
\end{enumerate}

To alleviate the emotion class imbalance issue, we utilize a weighted balanced data sampler to make sure samples from less representative categories have larger weights. 
In addition, within each training batch, we ensure the number of samples is balanced between the source and target domains.
As we mentioned before, in a real usage scenario, the amount of labeled data of the target domain is often limited. 
Considering this situation, domain adaptation is performed with a very limited number of labeled target samples under our experimental setting.
For samples of the target domain, we use only 10\% as the training set, and 40\% is used as a development set and the remaining is for evaluation. 

Considering the imbalance across different emotions, we report the average recall value, which is obtained by computing the per-class classification accuracy independently and then averaging over 4 different emotion categories.
For each experiment, we perform the experiments five times with exclusive training samples, and report the average 4-class emotion recognition recall value. As a comparison, we also conduct the following experiments:

\textbf{Source only}: the model is trained using source domain data only, and using the target domain data for evaluation.

\textbf{Source + Target}: 
The model is trained using samples both from source and target domains without any adversarial techniques.

The configuration of our model is shown in Table~\ref{tab:config}.

\begin{table}[h]
\vspace{-0.2cm}
\centering
\scalebox{0.75}{
\begin{tabular}{c|l}
\hline
\begin{tabular}[c]{@{}c@{}}Training \\ details:\end{tabular} & \begin{tabular}[c]{@{}l@{}}Adam optimizer (lr=0.001) +  L2 regularization (weight=0.01) \\ batch size=32; epochs=100; $\tau=2$.\end{tabular}                                                                                                                                                                               \\ \hline
ENC                                                          & \begin{tabular}[c]{@{}l@{}}Acoustic part:\\Conv1D(in\_ch=41,out\_ch=64, kernel size=10, stride=2), PReLU\\ Conv1D(in\_ch=64,out\_ch=128, kernel size=5, stride=2), PReLU\\ GRU(in\_size=128, hidden\_size=128, num\_layers=1)\\ 
Visual part:\\
GRU(in\_size=512, hidden\_size=512, num\_layers=1)\\ 
Merge part: \\
Linear(in=128+512, out=128), PReLU, Dropout\\
Linear(in=128, out=128)\end{tabular} \\ \hline
EC                                                           & \begin{tabular}[c]{@{}l@{}}Linear(in=128, out=32) PReLU\\ Linear(in=32, out=10) PReLU\\ Linear(in=10, out=4)\end{tabular}                                                                                                                                                                                       \\ \hline
DC                                                           & \begin{tabular}[c]{@{}l@{}}Linear(in=128, out=32) PReLU\\ Linear(in=32, out=10) PReLU\\ Linear(in=10, out=2)\end{tabular}                                                                                                                                                                                        \\ \hline
\end{tabular}
}
\caption{Model structure and training configuration details}
\vspace{-0.3 cm}
\label{tab:config}
\end{table}

\vspace{-0.2cm}
\section{Results and Discussion}

\subsection{Using NI and OI as source domain}
\label{subsubsec:NI_OI}
We use the NI and OI data as the source domain, and TR or TI as the target domain. Compared with TR, samples in TI have same fixed lexical content, but are recorded during an improvised dialog.

\begin{table}[h]
\scalebox{0.88}{
\begin{tabular}{c|c|c}
\hline
                                                                                & \multicolumn{2}{c}{Source Domain $\rightarrow$ Target Domain} \\ \hline
Experiment                                                                       &     NI+OI $\rightarrow$ TR   &       NI+OI $\rightarrow$ TI              \\ \hline
Source only                                                                       &     0.5005 (0.021)                           &         \textbf{0.7330} (0.020)                      \\ 
Source + Target                                                                   &        0.5837 (0.031)                        &          0.7288 (0.034)                     \\ 
Domain adversarial loss only                                                      &       0.6225 (0.018)                         &          0.7002 (0.007)                     \\ 
\begin{tabular}[c]{@{}c@{}}Domain adversarial loss\\ + softlabel loss\end{tabular} &     \textbf{0.6339} (0.023)                           &          0.7092 (0.022)                     \\ \hline
\end{tabular}}
\caption{Mean (std) of 4-class emotion recognition average recall value using NI+OI as source domain} \label{tab:PS}
\vspace{-0.5 cm}
\end{table}

The experiment results are shown in Table~\ref{tab:PS}. We first notice the domain mismatch exists between NI+OI and TR: the performance of ``source only" model has the lowest performance among all experiments. 
The ``Source + Target" experiment shows that adding target domain training samples helps in model domain adaptation as expected.
Even though only a few samples (10\% of the samples from the target domain) are employed during training, the performance improves from 0.5005 to 0.5837 directly. 
More importantly, the results of our proposed model show that the domain adversarial loss can help the target domain adaptation. The softlabel loss, considering the relationships across different emotion categories, also contributes to the domain adaptation under limited target domain samples with the highest performance of 0.6339.

However, in the experiment of NI+OI $\rightarrow$ TI, there is no improvement by adding target training samples. Surprisingly, in contrast to the previous findings, the domain adversarial loss and softlabel loss even worsen the performance. 
A closer inspection reveals that many samples within OI are exactly the same scripted utterances spoken in an improvised way as those in TI. 
As described in Table \ref{tab:msp_utt}, OI and TI were both recorded during the improvisation sessions. 
We find the actors often practiced for a while and repeated scripted utterances multiple times before they were ready for the Target-improvised (TI) recording in many improvisation sessions.
During dataset creation, those practicing utterances were largely kept and included in the group of OI. 

Thus,  OI utterances overlap with TI utterances in terms of the emotion elicitation manner and content information, which might consequently lead to the fact that adding new samples from target domain doesn't provide extra information of target domain. Moreover, this indistinct boundary between source and target can potentially lead to the domain confusion loss back propagating unintended bias during adaptation training, which can lead to a lower performance after adding the adversarial loss.
In addition, we observe a large absolute increase of accuracy from using TR to TI as target domain (from 0.5005 to 0.7330), which is likely caused due to the existing overlap between OI and TI.

Apart from this domain overlapping issue caused by the data collection process, the results from our experiment verify the first proposed hypothesis: improvised utterances can be used as source domain to help the emotion recognition for the utterances with the scripted utterances.

\vspace{-0.2cm}

\subsection{Using TR and TI as source domain}
We further reverse the source and target domain: TR and TI are used as the source domain and we select NI or OI as target domain. 
Under this setting, we are investigating whether we can apply the domain transfer learning from the acted dataset with fixed lexical content to those in natural conversation with variable lexical content.
Compared with the combinations in Sec.~\ref{subsubsec:NI_OI}, we have fewer samples in the source domain, but more samples in the target domain. 

\begin{table}[h]
\scalebox{0.88}{
\begin{tabular}{c|c|c}
\hline
                                                                                & \multicolumn{2}{c}{Source Domain $\rightarrow$ Target Domain} \\ \hline
Experiment                                                                       &     TI+TR $\rightarrow$ NI   &       TI+TR $\rightarrow$ OI              \\ \hline
Source only                                                                       &      0.4296(0.028)                           &         0.5499 (0.010)                      \\ 
Source + Target                                                                   &        0.4440 (0.032)                        &          0.5886 (0.010)                     \\ 
Domain adversarial loss only                                                      &       0.4601 (0.013)                         &          0.5904 (0.024)                     \\ 
\begin{tabular}[c]{@{}c@{}}Domain adversarial loss\\ + softlabel loss\end{tabular} &     \textbf{0.4653} (0.023)                           &          \textbf{0.5920} (0.017)                     \\ \hline
\end{tabular}}
\caption{Mean (std) of 4-class emotion recognition average recall value using TI+TR as source domain} \label{tab:TR}
\vspace{-0.6cm}
\end{table}

From Table \ref{tab:TR}, similarly, we notice the domain mismatch between source and target from the lower performance of ``Source only" models. 
A better emotion classification accuracy of ``Source+Target" model indicates that extra target training samples can help the model adaptation for the target domain.
For both experiments, we observe the domain adversarial loss and softlabel loss can help the model to achieve better performance with limited training samples of the target domain.
By comparing the results across two target domains, we notice NI has a lower classification accuracy. 
A reasonable explanation is that NI has a larger domain mismatch since both emotion elicitation manners and content information are different from the source.

All results show that we can improve the emotion recognition for daily natural conversation by leveraging domain adversarial loss and softlabel loss on the content fixed corpus. The results verify the hypotheses we proposed in the beginning.
\vspace{-0.2cm}
\section{Conclusion}

The goal of this work is to focus on the domain adaptation across emotion utterances under different elicitation approaches. 
We proposed two hypotheses and investigated transfer learning between utterances with fixed scripted lexical content and utterances collected from spontaneous, improvised production, particularly with limited labeled target samples.
Our adversarial training framework extracts domain invariant emotion representations. Moreover, we utilize the softlabel loss to take the correlations across different emotion categories into consideration.
Our results confirm our proposed hypotheses and further provide new insights into elicitation approaches in emotion data collection, as well as the importance of domain adaptation in emotion recognition.

\vfill\pagebreak

\bibliographystyle{IEEEtran}

\bibliography{mybib}

\end{document}